\documentclass[12pt]{article}
\usepackage[T1]{fontenc}
\usepackage{amsfonts}
\usepackage{latexsym}
\usepackage{dsfont}
\usepackage{amsmath}
\usepackage{amssymb}
\usepackage{amssymb}
\usepackage{amsthm}
\usepackage{mathrsfs}
\usepackage[nosort]{cite}
\usepackage[bulletsep]{collref}
\usepackage{setspace}
\usepackage{color}
\usepackage{graphicx}
\usepackage{accents}
\usepackage[font=small,labelfont=bf]{caption}
\usepackage[utf8]{inputenc}
\usepackage{geometry}
\usepackage[toc,page]{appendix}

\usepackage[linktocpage=true]{hyperref}
\hypersetup{
colorlinks=false,
citecolor= blue,
linkcolor= blue,
urlcolor= blue,
breaklinks=true
}
\usepackage{cleveref}
\crefformat{section}{\S#2#1#3} 
\crefformat{subsection}{\S#2#1#3}
\crefformat{subsubsection}{\S#2#1#3}

\def\appendix#1{\addtocounter{section}{1}\setcounter{equation}{0}
\renewcommand{\thesection}{\Alph{section}}
\section*{Appendix \thesection\protect\indent \parbox[t]{11.15cm}{#1}}
\addcontentsline{toc}{section}{Appendix \thesection\ \ \ #1}}

\allowdisplaybreaks
\numberwithin{equation}{section}

\makeatletter
 \let\old@startsection=\@startsection
 \let\oldl@section=\l@section
 \renewcommand{\@startsection}[6]{\old@startsection{#1}{#2}{#3}{#4}{#5}{#6\mathversion{bold}}}
 \renewcommand{\l@section}[2]{\oldl@section{\mathversion{bold}#1}{#2}}
\makeatother

\def\dd{\text{d}}
\def\lp{\lambda_+}
\def\lm{\lambda_-}
\def\ad{\text{ad}}
\def\L{\mathscr{L}}

\newtheorem{proposition}{Proposition}
\newtheorem{theorem}{Theorem}
\newtheorem{corollary}{Corollary}

\DeclareMathOperator*{\Tr}{Tr}

\begin{document}


\begin{titlepage}
\begin{center}
\vspace*{-1.0cm}
\hfill {\footnotesize NORDITA 2022-079}

\vspace{2.0cm}

{\LARGE  {\fontfamily{lmodern}\selectfont \bf Non-relativistic string monodromies}} \\[.2cm]

\vskip 1.5cm
\textsc{Andrea Fontanella$^{a}$, Juan Miguel Nieto Garc\'ia$^{b}$ \footnotesize
and \normalsize Olof Ohlsson Sax$^{c}$}\\
\vskip 1.2cm

\begin{small}
{}$^a$ \textit{Perimeter Institute for Theoretical Physics, \\
Waterloo, Ontario, N2L 2Y5, Canada} \\
\vspace{1mm}
\href{mailto:afontanella@perimeterinstitute.ca}{\texttt{afontanella@perimeterinstitute.ca}}

\vspace{5mm}
{}$^b$ \textit{II. Institut für Theoretische Physik, Universität Hamburg,\\
Luruper Chaussee 149, 22761 Hamburg, Germany} \\
\href{mailto:juan.miguel.nieto.garcia@desy.de}{\texttt{juan.miguel.nieto.garcia@desy.de}}

\vspace{5mm}
{}$^c$ \textit{Nordita,
Stockholm University and KTH Royal Institute of Technology\\
Hannes Alfvéns väg 12, SE-106 91 Stockholm, Sweden} \\
\href{mailto:Olof's email}{\texttt{olof.ohlsson.sax@nordita.org}}
\end{small}

\end{center}

\vskip 1 cm
\begin{abstract}
\vskip1cm\noindent
Spectral curve methods proved to be powerful techniques in the context of relativistic integrable string theories, since they allow to derive the semiclassical spectrum from the minimal knowledge of a Lax pair and a classical string solution. In this paper we initiate the study of the spectral curve for non-relativistic strings in AdS$_5\times S^5$. First we show that for string solutions whose Lax connection is independent of $\sigma$, the eigenvalues of the monodromy matrix do not have any spectral parameter dependence. We remark that this particular behaviour also appears for relativistic strings in flat space.  
Second, for some simple non-relativistic string solutions where the path ordered exponential of the Lax connection can be computed, we show that the monodromy matrix is either diagonalisable with quasi-momenta independent of the spectral parameter, or non-diagonalisable. For the latter case, we propose a notion of generalised quasi-momenta, based on maximal abelian subalgebras, which retain a dependence on the spectral parameter.  
\end{abstract}

\end{titlepage}

\tableofcontents
\vspace{5mm}
\hrule


\setcounter{section}{0}
\setcounter{footnote}{0}

\section*{Introduction}

The purpose of this paper is to initiate the application of analytic methods to identify the spectrum of non-relativistic (NR) strings in AdS$_5\times S^5$. The theory has been found in \cite{Gomis:2005pg}, which is a theory with non-relativistic target space but with a relativistic worldsheet. By taking a further NR limit in the worldsheet, the theory becomes Spin Matrix theory, a generalisation of the Landau-Lifshitz model \cite{Harmark:2017rpg,Harmark:2018cdl,Harmark:2019upf}. NR string theory has first been discovered in flat space \cite{Gomis:2000bd, Danielsson:2000gi}, and generalised in \cite{Bergshoeff:2018yvt} for generic curved backgrounds. The ambient geometry seen by a NR string is non-Lorentzian, and identified to be a string Newton-Cartan geometry.  NR string theory is free of Weyl anomalies, and imposing the vanishing of the beta function implies the background fields need to fulfil some NR (super)gravity equations \cite{Gomis:2019zyu}. There are two main approaches to derive a theory of NR strings. One is the limit approach \cite{Gomis:2005pg, Gomis:2000bd,Bergshoeff:2018yvt,Bergshoeff:2019pij}, while the second one is the null-reduction method \cite{Harmark:2017rpg,Harmark:2018cdl,Harmark:2019upf}, which derives a NR string action starting from a relativistic one by dimensionally reducing the target space along a null isometry direction. The two procedures have been proved to be equivalent under a technical assumption. There is also a third approach, based on the expansion of the string action at higher orders in the speed of light parameter \cite{Hartong:2021ekg,Hartong:2022dsx}. 
Many other topics have been studied, such as action symmetries \cite{Harmark:2019upf, Bergshoeff:2018yvt, Bergshoeff:2019pij, Bidussi:2021ujm}, connection to double field theory \cite{Ko:2015rha, Blair:2019qwi, Blair:2020gng, Blair:2020ops}, Hamiltonian formalism \cite{Kluson:2017abm,Kluson:2018egd,Kluson:2018grx}, and open strings \cite{Gomis:2020fui,Gomis:2020izd}.
For a recent review on NR strings, see \cite{Oling:2022fft} and references therein. 

Motivations to study NR strings can come from different angles, such as providing a different perspective to quantum gravity, seen as the UV completion of a NR theory of GR. For us the motivation comes from holography, as the target space of NR strings is a string Newton-Cartan manifold, and therefore they provide an interesting arena where to study non-Lorentzian holography.
Some progress has been made in understanding the string side of a NR AdS$_5$/CFT$_4$ correspondence, in particular regarding classical NR string solutions~\cite{Fontanella:2021btt}, semiclassical expansion~\cite{Fontanella:2021hcb}, coset space formulation of NR strings in AdS$_5\times S^5$ and Lax pair~\cite{Fontanella:2022fjd,Fontanella:2022pbm}. 
The ultimate goal is to understand the spectrum of NR strings in AdS$_5\times S^5$, to possibly match with some data, like scaling dimensions of operators, of the dual field theory which is still undetermined.  


Finding the spectrum of strings in a generic background is a very hard problem. However, in certain AdS backgrounds a lot of progress have been made by using the integrability of the worldsheet theories. In particular, in the case of relativistic AdS$_5 \times S^5$ the full quantum spectrum at large $N$ can be obtained using intergability.\footnote{For an overview of integrability in AdS$_5 \times S^5$ see the reviews~\cite{Arutyunov:2009ga,Beisert:2010jr}.} Similarly, the spectrum of strings in AdS$_4 \times CP^3$~\cite{Stefanski:2008ik,Arutyunov:2008if} and various AdS$_3 \times S^3 \times M_4$ backgrounds~\cite{Babichenko:2009dk,OhlssonSax:2011ms,Cagnazzo:2012se} have been studied using the integrability of the worldsheet theories.

At a classical level, integrability of these string backgrounds manifests itself through the reformulation of the equations of motion in terms of a Lax pair. Such a description is possible for any background which can be written as a symmetric coset space.\footnote{For supersymmetric backgrounds the corresponding coset is a semi-symmetric space.} Using the Lax pair we can construct a \emph{monodromy matrix} -- the integral of the Lax connection around the worldsheet -- whose trace, the \emph{transfer matrix}, is independent of the worldsheet time $\tau$ and acts as a generating function for the higher conserved charges of the integrable model. The eigenvalues of the monodromy matrix span a Riemann surface, known as the \emph{spectral curve},\footnote{For a review of the spectral curve of AdS$_5 \times S^5$ see~\cite{SchaferNameki:2010jy}. A general classification of spectral curves for AdS backgrounds was given in~\cite{Zarembo:2010yz}.} which not only encodes the spectral data of the string solution, but also provides an efficient method for finding the leading semi-classical quantum corrections~\cite{Gromov:2008ec}. Understanding the spectral curve corresponding to a solution to the classical string equations of motion is an important step in the application of integrability based methods to string theory. In this paper we will initiate a study of the spectral curve of strings in the NR AdS$_5\times S^5$ background.

This paper is structured as follows. In section~\ref{sec:NR_string_setting} we give the setting of the paper. We introduce the NR string action in AdS$_5\times S^5$, review the coset description and the Lax pair. In section~\ref{sec:monodromy} we study solutions to the characteristic equation for the monodromy matrix, and present our main theorem. We discuss the monodromy matrix evaluated on two string solutions, leading to non-diagonalisable matrices. We propose in this case a generalisation of the notion of quasi-momenta, based on the maximal abelian subalgebra decomposition. In section~\ref{sec:flat_space} we discuss the monodromy matrix for relativistic strings in flat space, where we find again, as in the NR context, a non-diagonalisable monodromy matrix with eigenvalues independent of spectral parameter.

\section{NR string action in AdS$_5\times S^5$}\label{sec:NR_string_setting}
As first obtained in \cite{Gomis:2005pg}, the NR string action in AdS$_5\times S^5$ is\footnote{We refer to\cite{Fontanella:2021hcb} for other choices of coordinates.} 
\begin{equation}\label{NR_AdS_action}
\begin{aligned}
S &= - \frac{T}{2} \int \dd^2 \sigma \, \bigg[\gamma^{\alpha\beta}\bigg( x^a x_a (-\partial_{\alpha} x^0 \partial_{\beta} x^0 + \cos^2 x^0 \partial_{\alpha} x^1 \partial_{\beta} x^1)
+\partial_{\alpha} x^a \partial_{\beta} x_a + \partial_{\alpha} x^{a'} \partial_{\beta} x_{a'} \bigg)\\
&+ \varepsilon^{\alpha\beta} \bigg( (\lp e_{\alpha}{}^+ + \lm e_{\alpha}{}^-) \partial_{\beta} x^0 + (\lp e_{\alpha}{}^+ - \lm e_{\alpha}{}^-) \cos x^0 \partial_{\beta} x^1 \bigg) \bigg] \ ,
\end{aligned}
\end{equation}
where $T$ is the string tension, $\sigma^{\alpha} = (\tau, \sigma)$, with $\alpha = 0, 1$, are the string worldsheet coordinates, $\gamma^{\alpha\beta} \equiv \sqrt{-h} h^{\alpha\beta}$ is the Weyl invariant combination of the inverse worldsheet metric $h^{\alpha\beta}$ and $h =$ det$(h_{\alpha\beta})$, and $e_{\alpha}{}^{\pm}$ are (the light-cone components of) the worldsheet zweibein, see appendix \ref{app:Conventions} for our conventions. The $x^A \equiv (x^0, x^1)$ are longitudinal coordinates originating from AdS$_5$, while $x^a$ and $x^{a'}$, with $a, b, ... = 2, 3, 4$ and $a', b', ... = 5, ..., 9$, are transverse coordinates originating from AdS$_5$ and $S^5$ respectively, which are contracted with $\delta_{ab}$. $\lambda_{\pm}$ are non-dynamical (Lagrange multipliers) scalar fields, depending on the worldsheet coordinates.

\subsection{Closed string solutions}
\label{sec:string_solutions}
In this section we present two closed string solutions relevant for the spectral curve discussion. We start by fixing conformal gauge, i.e., $h_{\alpha\beta} = \eta_{\alpha\beta}$. The worldsheet zweibein $e_{\alpha}{}^{\pm}$ is fixed as in \cite{Fontanella:2021btt}, 
\begin{equation}
\label{zweibein_conformal_gauge}
e_{\alpha}{}^+ = (-1,-1) \, , \qquad\qquad
e_{\alpha}{}^- = (-1,1) \ .
\end{equation}
The closed string solution must satisfy the equations of motion for the target space coordinates and Lagrange multipliers, $\Phi \equiv (x^0, x^1, x^a,x^ {a'}, \lambda_{\pm})$, and the Virasoro constraints. On top, closed string boundary conditions are imposed, i.e., $\Phi(\tau, \sigma + 2\pi) = \Phi(\tau, \sigma)$. 

In conformal gauge, (a combination of) the equations of motion for the Lagrange multipliers gives 
\begin{equation}
\label{Lagrange_eom}
\begin{aligned}
\frac{1}{2 \cos x^0}(\mathcal{E}_{\lambda_+} + \mathcal{E}_{\lambda_-}) &=  -  x'^0 \sec x^0  +  \dot{x}^1   = 0 \, , \\
\frac{1}{2 \cos x^0}(\mathcal{E}_{\lambda_+} - \mathcal{E}_{\lambda_-}) &= \dot{x}^0 \sec x^0 -  x'^1   = 0 \, ,
\end{aligned}
\end{equation}
where dots and primes indicates $\partial_{\tau}$ and $\partial_{\sigma}$, respectively. By introducing the variable $y$ defined by integration of $\dd y = \sec (x^0)  \dd x^0$, the two equations above become 
\begin{equation}
- y' + \dot{x}^1 = \dot{y} - x'^1 = 0 \, , 
\end{equation}
which imply that $y$ and $x^1$ satisfy the wave equation, whose most general solution is
\begin{equation}
y = f_+ (\sigma_+) + f_- (\sigma_-) \, , \qquad\qquad 
x^1 =  f_+ (\sigma_+) - f_- (\sigma_-) + \text{const.} \, , 
\end{equation}
where $f_{\pm}$ are generic functions of their argument $\sigma_{\pm} \equiv \tau \pm \sigma$, which parametrise the Diff$_+\oplus$Diff$_-$ residual gauge symmetry after imposing conformal gauge. We may fix such redundancy by taking $f_{\pm} =$ Id, which gives
\begin{equation}
y = \tau \, , \qquad\qquad x^1 = \sigma \, ,
\end{equation}
where we also set the constant to zero. The solution $x^1 = \sigma$ is clearly not periodic in $\sigma$. We use the fact $x^1$ is an isometry for the background geometry of the NR action (\ref{NR_AdS_action}), and we make it compact $x^1 \sim x^1 + 2 \pi$. Under this assumption, $x^1 = \sigma$ describes a closed string. 
After this short introduction, we are ready to present the two string solutions considered in this paper. 
\begin{itemize}
\item \emph{Longitudinal solution}: This is the simplest solution (with non-trivial $x^0$) admitted by the theory, given by 
\begin{equation}\label{longitudinal_sol}
x^0 = \text{gd}(\tau) \, , \qquad
x^1 = \sigma \, , \qquad
\text{others} = 0  \, , 
\end{equation}
where $y$ has been inverted for $x^0$, and $\text{gd}(\cdot)$ is the Gudermannian function, which can be expressed in terms of trigonometric functions as $\text{gd}(x) = 2 \arctan \left(\tanh \frac{x}{2}\right)$. This solution has non-trivial dynamics only on the longitudinal coordinates. 
\item \emph{BMN-like solution}: This solution is reminiscent of the BMN solution, as it has the simplest non-trivial dynamical behaviour we can write in both the old AdS and sphere coordinates
\begin{equation}\label{BMN_like_sol}
x^0 = \text{gd}(\tau) \, , \qquad
x^1 = \sigma \, , \qquad
x^5 = J \tau \, , \qquad
\lambda_{\pm} = \pm \frac{J^2}{2} \cosh \tau \, , \qquad
\text{others} = 0 \, ,
\end{equation}
where $J$ is a constant with meaning of linear momentum, as the 5-sphere becomes $\mathbb{R}^5$ in the NR limit. In contrast to their relativistic counterparts, the Virasoro constraints tie the dynamics on the $x^5$ coordinate to the Lagrange multipliers $\lambda_\pm$ instead of tying them to $x^0$ and $x^1$.
\end{itemize}

\subsection{Coset description of the AdS$_5\times S^5$ NR string action}
As found in \cite{Fontanella:2022fjd}, and later generalised in \cite{Fontanella:2022pbm}, the coordinate NR action (\ref{NR_AdS_action}) admits a coset $G/H$ description. In this paper we shall work with the setting given in \cite{Fontanella:2022pbm}, since it brings two useful advantages, namely the Lie algebra $\mathfrak{g}$ of the group $G$ admits a spinorial representation and an adjoint invariant inner product under the full $\mathfrak{g}$. 
The Lie algebra $\mathfrak{g}$ of the group $G$ is chosen to be the direct sum of the extended string Newton-Hooke and extended Euclidean algebras. The extensions are given by the NR Lie algebra expansion \cite{Fontanella:2020eje} applied to the AdS$_5\times S^5$ isometry algebra.  

The extended string Newton-Hooke algebra is spanned by a longitudinal boost $M$, longitudinal translations $H_A$, transverse rotations $J_{ab}$, transverse translations $P_a$, string-Galilei boosts $G_{Ab}$, and non-central extensions $Z_A, Z_{ab}$ and $Z$, with commutation relations  
\begin{equation} \label{ext_NH_algebra}
	\begin{aligned}
	{}[P_a, P_b] &= Z_{ab} \, , &[H_A, P_b] &= G_{Ab} \, , \\
	[J_{ab}, J_{cd}] & = \delta_{bc} J_{ad} - \delta_{ac} J_{bd} + \delta_{ad} J_{bc} - \delta_{bd} J_{ac} \,, \qquad &  [M,G_{Aa}]  & = -\varepsilon_A{}^B G_{Ba}\,, \\
    [J_{ab},P_{c}] & =  \delta_{bc} P_{a} - \delta_{ac} P_{b}\,, &[H_A,Z_B] & = -\varepsilon_{AB} Z \,, \\
    [J_{ab},G_{Ac}] & = \delta_{bc} G_{Aa} - \delta_{ac} G_{Ab} \,,& [M,Z_A] & = -\varepsilon_A{}^B Z_B\,, \\
	[G_{Aa}, G_{Bb}] & = \delta_{ab} \varepsilon_{AB} Z\,, & [Z,H_A]  & = -\varepsilon_A{}^B Z_B\,,\\
	[G_{Aa}, P_{b}] & = \delta_{ab} Z_A\,,& [M,H_A] & = -\varepsilon_A{}^B H_B\,, \\
	[G_{Aa},H_B]  & = -\eta^{}_{AB} P_{a}\,, 
	&[G_{Aa}, G_{Bb}] & = \delta_{ab} \varepsilon_{AB} Z - \eta_{AB} Z_{ab} \, , \\
	[J_{ab}, Z_{cd}] & = \delta_{bc} Z_{ad} - \delta_{ac} Z_{bd} + \delta_{ad} Z_{bc} - \delta_{bd}
 Z_{ac} \, ,  
 &[H_A, H_B]& = -\varepsilon_{AB} M \, ,
	\end{aligned}
\end{equation}
The extended Euclidean algebra is spanned by spatial translations $P_{a'}$, spatial rotations $J_{a'b'}$ and non-central extensions $Z_{a'b'}$, with commutation relations
\begin{equation}
\label{Euclidean}
\begin{aligned}
{}[P_{a'}, P_{b'}] &= - Z_{a'b'} \, , \\
[P_{a'}, J_{b'c'}] &= \delta_{a'b'} P_{c'} - \delta_{a'c'} P_{b'} \, , \\
[J_{a'b'}, J_{c'd'}] &= \delta_{b'c'} J_{a'd'} - \delta_{a'c'} J_{b'd'} + \delta_{a'd'} J_{b'c'} - \delta_{b'd'} J_{a'c'} \, , \\
 [J_{a'b'}, Z_{c'd'}] &= \delta_{b'c'} Z_{a'd'} - \delta_{a'c'} Z_{b'd'} + \delta_{a'd'} Z_{b'c'} - \delta_{b'd'} Z_{a'c'} \, .
\end{aligned}
\end{equation}
The gauge group $H$, with Lie algebra $\mathfrak{h}$, is taken to be everything that generates $G$ except of $H_A, P_a, P_{a'}$. The algebra $\mathfrak{g}$ admits a decomposition under a $\mathbb{Z}_2$ automorphism, $\mathfrak{g} = \mathfrak{g}^{(0)} \oplus \mathfrak{g}^{(1)} $, where 
\begin{equation}
\label{eq:grading}
\mathfrak{g}^{(0)} = \text{span} \{ M, J_{ab}, G_{Aa}, Z_{ab}, Z, J_{a'b'}, Z_{a'b'} \}\ , \qquad
\mathfrak{g}^{(1)} = \text{span} \{ H_A, P_a, P_{a'}, Z_A \}  \ ,
\end{equation}
where $\mathfrak{g}^{(0)}$ and $\mathfrak{g}^{(1)}$ are eigenspaces with eigenvalues $1$ or $-1$ under the action of the $\mathbb{Z}_2$ automorphism, respectively. We denote by $\mathbb P$ the projector into $\mathfrak{g}^{(1)}$. 

By taking a group element $g \in G$, we construct the Maurer-Cartan (MC) 1-form $A = g^{-1} \dd g$ whose components are
\begin{equation} 
	\begin{aligned}
A &= A^{H_A} H_A + A^{P_a} P_a + A^{P_{a'}} P_{a'}+ A^{Z_A} Z_A + A^{M} M + \frac{1}{2} A^{J_{ab}} J_{ab}  \\
& + A^{G_{Aa}} G_{Aa} + \frac{1}{2} A^{Z_{ab}} Z_{ab}+ A^{Z} Z + \frac{1}{2} A^{J_{a'b'}} J_{a'b'} + \frac{1}{2} A^{Z_{a'b'}} Z_{a'b'} \ ,
	\end{aligned}
\end{equation}
The Lagrange multipliers are collected into a separate algebra-valued 1-form $\Lambda$, 
\begin{equation}
\Lambda_{\alpha} = \Lambda_{\alpha}^{Z_A}\, Z_A  \equiv \lm e_{\alpha}{}^-  \, Z_+ + \lp e_{\alpha}{}^+  \, Z_- \ ,
\end{equation}
which combines with $A$ into the \emph{generalised} MC current $J_{\alpha}$, defined as 
\begin{equation}\label{gen_MC}
J_{\alpha} \equiv A_{\alpha} - (\star \Lambda)_{\alpha} = A_{\alpha} + \gamma_{\alpha\beta}\varepsilon^{\beta\gamma}\Lambda_{\gamma} \ .
\end{equation}
The coordinate action (\ref{NR_AdS_action}) then can be written in a coordinate-free language as
\begin{equation}
\label{NR_coset_action}
S^{G/H} = - \frac{T}{2} \int \dd^2 \sigma \, \gamma^{\alpha\beta}  \langle J^{(1)}_{\alpha}, J^{(1)}_{\beta} \rangle \ , \qquad\qquad
J^{(1)} \equiv {\mathbb P} J
\end{equation}
where $\langle \cdot, \cdot \rangle$ is an inner product on $\mathfrak{g}$ invariant under the full adjoint action of $\mathfrak{g}$, given in \cite{Fontanella:2022pbm}, chosen such that $\langle H_+, H_- \rangle = 0$. Such inner product is grading compatible, i.e., non-vanishing only when both elements have same grading.  

\subsection{Comment on the choice of coordinates}
\label{sec:coords}
We discuss here the issue of implementing a prescribed coset representative in the coordinate-free relativistic and NR AdS$_5\times S^5$ string actions. There are two procedures that one can follow to derive the NR string action in AdS$_5\times S^5$:
\begin{itemize}
\item \emph{Limit procedure}. We take a coset representative and plug it into the relativistic AdS$_5\times S^5$ action, rescale coordinates by the contraction parameter $c$ in the way suggested by the contraction of the isometry algebra of AdS$_5\times S^5$ to string Newton-Hooke (without extensions here), and take the limit on $c$. This will lead to a NR string action written in a specific set of coordinates, as done originally in \cite{Gomis:2005pg} for the ``GGK'' coordinates, or as generalised in \cite{Fontanella:2021hcb} for cartesian and polar global coordinates.
\item \emph{Algebraic procedure}. We take a coset representative (the same that has been taken in the limit procedure), but assuming that the generators satisfy the \emph{extended} string Newton-Hooke and Euclidean algebras (\ref{ext_NH_algebra}), (\ref{Euclidean}), with the generator identification $H_A \equiv P_A$, $G_{Ab} \equiv J_{Ab}$, and we plug it into the NR string action (\ref{NR_coset_action}).  
\end{itemize}

One may wonder whether the two procedures give the same action. Before answering that, we have to keep in mind that different parametrisations of the same space lead to different non-relativistic limits, as discussed in Appendix A of \cite{Fontanella:2021hcb}. Three particularly interesting coset representatives are

\vspace{2mm}
\noindent \emph{GGK} \hspace{4cm}$g = e^{x^1 P_1} e^{x^0 P_0} e^{x^a P_a} e^{x^{a'} P_{a'}}$,  

\vspace{4mm}
\noindent\emph{Polar} \hspace{2cm}$g = e^{t P_0 - \psi_1 J_{12} - \psi_2 J_{34}} e^{\arcsin x J_{13}} e^{\text{arcsinh} \rho P_1}
e^{\phi P_9 - \chi_1 J_{56} - \chi_2 J_{78}} e^{\arcsin w J_{57}} e^{\arcsin r P_5}$,

\vspace{4mm}
\noindent\emph{Cartesian} \hspace{3cm}$g = e^{t P_0} e^{z_i P_i} e^{\phi P_5} e^{y_i P_{5+i}} \, , \qquad\qquad i = 1,... , 4$,
\vspace{2mm}

\noindent What we found by direct computation is that, among the coset representatives given above, only GGK and polar coordinates\footnote{In polar coordinates the results are the same if in the limit procedure one rescales generators leaving coordinates as they are. If one wants to rescale coordinates, and leave generators untouched, then the two actions match if one linearises $\arcsin x$ and $\arcsin r$ with $x$ and $r$ respectively in the coset representative for the algebraic procedure.} give the same result in the two procedures. For the cartesian coordinates, the two methods do not give the same result, and we did not find a way to modify the coset representative for the algebraic procedure to get a match.

\subsection{Lax pair and monodromy matrix}

The independent, and some of the redundant\footnote{Some of the equations of motion obtained by varying $g$ in (\ref{NR_coset_action}) are redundant, and gauge invariance relates them via Noether identities. In the end, the number of independent equations of motion matches the coset degrees of freedom.}, equations of motion obtained from (\ref{NR_coset_action}) can be written as
\begin{equation}
\label{eom_compact}
\partial_{\alpha} (\gamma^{\alpha\beta} J^{(1)}_\beta) + \gamma^{\alpha\beta} [J_{\alpha}, J^{(1)}_\beta] = 0 \, ,
\end{equation}
where $J_{\alpha}$ is the generalised current defined in \ref{gen_MC}.
However, the equations of motion for the Lagrange multipliers, 
\begin{equation}
 \label{Lagrange_coset_eom}
\mathcal{E}^{\lp} \equiv  \varepsilon^{\alpha\beta} e_{\alpha}{}^+ A_{\beta}^{H_+} = 0 \ , \qquad\qquad
\mathcal{E}^{\lm} \equiv \varepsilon^{\alpha\beta}e_{\alpha}{}^- A_{\beta}^{H_-} = 0 \  .
\end{equation}
cannot be incorporated into (\ref{eom_compact}), which instead will be imposed as a constraint. The equations of motion written in the form (\ref{eom_compact}) admit a representation in terms of the Lax connection 
\begin{equation}
\label{Lax}
\L_\alpha = A_\alpha^{(0)} + \frac{z^2 +1}{z^2-1} A_\alpha^{(1)} - \frac{2z}{z^2-1} \gamma_{\alpha\beta}\varepsilon^{\beta \gamma} J_\gamma^{(1)}.
\end{equation}
where $z$ is the spectral parameter. The Lax representation of the equations of motion holds on solutions of the constraint equations (\ref{Lagrange_coset_eom}).

It is well known that a Lax pair codifies information about an infinite tower of conserved quantities of the dynamical system it is associated to. In order to extract them, we first have to construct the monodromy matrix $\cal M$
\begin{equation}
    {\cal M} (z,\tau)= {\cal P}\!\exp \left[ \int_0^{2\pi} \L_\sigma \, \dd\sigma \right] \ , \label{monodromymatrix}
\end{equation}
where ${\cal P}\!\exp$ indicates that the integrals in the exponential are path-ordered. Flatness of the Lax pair guarantees we can choose to perform the integral just along the $\sigma$ direction. In addition, it also imples that the monodromy matrix satisfies the following evolution equation
\begin{equation}
    \partial_{\tau} \mathcal{M} =\L_\tau(z  , \tau, 2\pi) \mathcal{M} - \mathcal{M} \L_\tau(z , \tau, 0)=[\L_\tau(z , \tau, 0) , \mathcal{M}] \ , \label{evolution}
\end{equation}
where we have specified the explicit dependence of the $\tau$ component of the Lax pair on the spectral parameter $z$ and the worldsheet coordinates. In the last equality we have assumed that $\L_\tau$ is periodic in the worldsheet $\sigma$ coordinate. It is now trivial to check that the trace of powers of the monodromy matrix, $\mathcal{H}_k=\Tr [\mathcal{M}^k]$, are constants of motion.\footnote{Checking if these constants of motion are in involution would require us to study the Poisson structure of the Lax connection, which is outside the scope of this article}.

Among the methods at our disposal to understand and solve integrable models, one of them is to study the locus of zeros of the characteristic polynomial of the monodromy matrix, i.e., the Riemann surface defined by the eigenvalues of the monodromy matrix as functions of the spectral parameter
\begin{equation}
\det [\mathcal{M} (z,\tau)- \omega (z) \,\text{Id}] = 0 \ .
\end{equation}
The eigenvalues $\omega$ cannot depend on $\tau$, as they can be expressed as linear combinations of $\mathcal{H}_k$.

For later convenience, we define the quasi-momenta $p(z)$ as $\omega =e^{i p}$. In the literature regarding strings on AdS$_5 \times S^5$, quasi-momenta associated to the AdS space are usually denoted with a hat, $\hat{p}$, while the ones associated to the sphere are denoted with a tilde, $\tilde{p}$. We will use the same notation for the non-relativistic background.

Similarly to relativistic AdS$_5 \times S^5$, our Lax connection becomes the Maurer-Cartan current in the $z\rightarrow \infty$ limit. This means that
\begin{equation}\label{rep_free_asympt_monodromy}
    \lim_{z\rightarrow \infty} \mathcal{M} (z,\tau)={\cal P}\!\exp \left[ \int_0^{2\pi} A_\sigma \, \dd\sigma \right] = {\cal P}\!\exp \left[ \int_0^{2\pi} g^{-1} \partial_\sigma g \, \dd\sigma \right] = g(\tau, 2\pi) g(\tau, 0 )^{-1} \ .
\end{equation}
Due to the periodicity of our solutions, this imposes $\lim_{z\rightarrow \infty} \omega (z)=1$. However, this does not imply that the quasimomenta vanish in this limit. In fact, $\lim_{z\rightarrow \infty} p(z)= 2\pi m$, where $m$ carries information regarding winding numbers.

\section{Some observations on the monodromy matrix}\label{sec:monodromy}

The goal is to write down the spectral curve associated to a NR string solution and extract the spectrum. The first step is to compute the eigenvalues of the monodromy matrix when evaluated on a given string solution. However, as we are going to explain,  we find a universal behaviour of the monodromy matrix eigenvalues, which holds regardless of the particular string solution considered. The eigenvalues will be computed using two different representations: one is the spinorial representation obtained by Lie algebra expansion from the relativistic isometry algebra \cite{Fontanella:2022pbm}, the second one is the adjoint representation for the ``less'' extended algebra as initially given in \cite{Fontanella:2022fjd}, see Appendix \ref{app:Rep} for the detail. 
In both representations, although the degeneracy of eigenvalues is different, the physical result is the same. Our result will be presented in the spinorial representation.

\subsection{Quasi-momenta are spectral parameter independent}

\begin{proposition}\label{theorem1}
On solutions of the constraint (\ref{Lagrange_coset_eom}), the eigenvalues of $\L_{\sigma}$ do not depend on the spectral parameter.
\end{proposition}

\begin{proof}
We take the GGK set of coordinates given by\footnote{The proof also holds for polar coordinates, with the modification of the coset representative as discussed in section \ref{sec:coords}.}
\begin{equation}
g = e^{x^1 H_1} e^{x^0 H_0} e^{x^a P_a} e^{x^{a'} P_{a'}} \, ,  
\end{equation}
and we evaluate the MC 1-form by using the spinorial representation, see Appendix \ref{app:Rep}. Then we compute the Lax pair (\ref{Lax}), but we do not evaluate it on any string solution, so the result will be independent of the solution chosen. For convenience, we fix conformal gauge $h_{\alpha\beta} = \eta_{\alpha\beta}$ and we fix the zweiben as in (\ref{zweibein_conformal_gauge}). The Lax matrix in the spinorial representation is a 24\texttimes 24 matrix with two blocks: the 12\texttimes 12 AdS$_5$ block and the 12\texttimes 12 $S^5$ block.   

The characteristic equations for the $\sigma$-component of the Lax connection,
\begin{equation}
\det (\L_{\sigma} - \mu \,\text{Id}) = 0  
\end{equation} 
admits the following solutions, where $\hat{\mu}$ and $\tilde{\mu}$ belong to the AdS$_5$ and S$^5$ blocks respectively. For the AdS$_5$ block we have
\begin{equation}
\begin{aligned}
\hat{\mu}^{\pm}_1 &= \pm \sqrt{a + b + 2 \sqrt{ab}} \, , & \qquad\qquad\text{with multiplicity  3} \, , \\
\hat{\mu}^{\pm}_2 &= \pm \sqrt{a + b - 2 \sqrt{ab}} \, , &\qquad\qquad\text{with multiplicity  3} \, , \\
\end{aligned}
\end{equation}
where we defined 
\begin{equation}
a \equiv -\L^{H_+}_{\sigma} \L^{H_-}_{\sigma} + (\L^{M}_{\sigma})^2 \, ,  \qquad\quad
b \equiv -(\L^{J_{23}}_{\sigma})^2-(\L^{J_{24}}_{\sigma})^2-(\L^{J_{34}}_{\sigma})^2 \, .
\end{equation}
For the $S^5$ block we get
\begin{equation}
\begin{aligned}
\tilde{\mu}_i &=  \, g_i (\L^{J_{a'b'}}_{\sigma})  \, , &\qquad\qquad i = 1, ..., 12 \, , 
\end{aligned}
\end{equation}
where $g_i$ are functions of the $\L^{J_{a'b'}}_{\sigma}$ components only, which we do not need to spell out as they do not carry any spectral parameter by definition.  
The only contributions in $\hat{\mu}^{\pm}_1, \hat{\mu}^{\pm}_2$ which may bring a dependence on the spectral parameter must be coming from $\L^{H_+}_{\sigma}$ and $\L^{H_-}_{\sigma}$, since all other terms belong to $\mathfrak{g}^{(0)}$, so do not carry any dependence on $z$. These two terms are 
\begin{equation}
\L^{H_{\pm}}_{\sigma} = \frac{z^2+1}{z^2-1} \left( x'^0 \pm \cos x^0 x'^1 \right) 
+ \frac{2 z}{z^2-1} \left( \dot{x}^0 \pm \cos x^0 \dot{x}^1 \right)  \, ,
\end{equation}
which only appear in the quasi-momenta via the product $\L^{H_+}_{\sigma} \L^{H_-}_{\sigma}$. When such product is evaluated on solutions of (\ref{Lagrange_coset_eom}), it becomes 
\begin{equation}
\L^{H_+}_{\sigma} \L^{H_-}_{\sigma} = \cos^2 x^0 \left[ (\dot{x}^1)^2 - (x'^1)^2\right] \, .
\end{equation}
This ends the proof, since the result is independent of the spectral parameter $z$.
\end{proof}

\begin{theorem}\label{corollary1}
On solutions of the constraint (\ref{Lagrange_coset_eom}) and for $\L_{\sigma}$ independent of $\sigma$, the eigenvalues of the monodromy matrix do not depend on the spectral parameter.
\end{theorem}

\begin{proof}
Because $\L_{\sigma}$ does not depend on $\sigma$, the path ordered exponential in~\eqref{monodromymatrix} simplifies drastically to 
\begin{equation}
\mathcal{M}(z) =  e^{2\pi \L_{\sigma}(z)} \, . 
\end{equation}
Then the eigenvalues for the monodromy matrix are related to the eigenvalues of $\L_{\sigma}$ by\footnote{If $\L_{\sigma}$ is diagonalisable, then this is straightforward, but if it is non-diagonalisable, then one can prove this relation by transforming $\L_{\sigma}$ in its Jordan normal form. } 
\begin{equation}
\omega = e^{2 \pi \mu} 
\qquad\text{or}\qquad
 p=-2\pi i \mu \, . 
\end{equation}
Since on shell $\mu$ does not depend on $z$, neither $\omega$ nor $p$ depend on $z$.  
\end{proof}

\begin{corollary}
On solutions of the constraint (\ref{Lagrange_coset_eom}) and for $\L_{\sigma}$ independent of $\sigma$, the quasimomenta $p$ are proportional to the winding numbers of the solutions.
\end{corollary}

\begin{proof}
Theorem~\ref{corollary1} guarantees that the quasimomenta are independent of the spectral parameter. Combining this result with the asymptotic limit of the monodromy matrix we discussed in the previous section, $\lim_{z\rightarrow \infty} \mathcal{M} (z,\tau)= g(\tau, 2\pi) g(\tau, 0)^{-1}$, the corollary follows immediately.
\end{proof}


\subsection{Diagonalisable monodromy matrix}

Suppose that $\L_{\sigma}$ is $\sigma$ independent, so that its path ordered exponential can be computed, and suppose the resulting monodromy matrix is diagonalisable. This is the case, e.g., when the Lax connection is evaluated on the longitudinal solution (\ref{longitudinal_sol}). 
Then the monodromy matrix can be written as
\begin{equation}\label{M_diagonal}
\mathcal{M} = U e^{i p_i C_i} U^{-1} \, ,  
\end{equation}
where $C_i$ are generators of the Cartan subalgebra (see Appendix \ref{app:Rep}), $p_i$ are the quasi-momenta and $U$ is the similarity transformation that brings $\mathcal{M}$ into the diagonal form. As a direct consequence of the theorem~\ref{corollary1}, the quasi-momenta do not depend on the spectral parameter. This hinders any application of the spectral curve method, e.g., the ones described in \cite{Gromov:2007aq,Gromov:2007cd,Gromov:2008ie,Gromov:2007aq}, since the analytic structure of the quasi-momenta on the spectral parameter is trivial.

\subsection{Non-diagonalisable monodromy matrix}

Suppose again that $\L_{\sigma}$ is $\sigma$ independent, but the monodromy matrix is non-diagonalisable. This scenario is realised when the Lax connection is evaluated, e.g., on the BMN-like solution (\ref{BMN_like_sol}).

In this case the monodromy matrix cannot be brought into the diagonal form (\ref{M_diagonal}), however one can find a similarity transformation $S$ such that $\mathcal{M}$ takes the Jordan normal form. It is tempting to guess that the Jordan normal form of $\mathcal{M}$ may be realised by the \emph{maximal abelian subalgebra} (MAS) instead of the Cartan subalgebra\footnote{This is motivated by the fact that the Cartan subalgebra is defined by the set of generators which commute among themselves and whose adjoint representation is diagonalisable. Our generalisation consists in dropping the diagonalisability requirement. However, although the Cartan subalgebra is not unique, it can only have a fixed dimension, while the MAS is not unique and it may differ in dimension.}, such that the monodromy matrix may be written as
\begin{equation}\label{M_non_diag}
\mathcal{M} = S e^{i q_i W_i} S^{-1} \, , 
\end{equation} 
where $W_i$ are generators of the MAS and $q_i$ are generalised quasi-momenta. In this case, theorem~\ref{corollary1} would only apply to the quasimomenta associated to the Cartan subalgebra of the MAS we are considering. The remaining quasimomenta can, in principle, depend on the spectral parameter.

As a concrete example, we consider the monodromy matrix evaluated on the BMN-like solution,
\begin{equation}\label{M_BMN_like}
\mathcal{M} = e^{2\pi (a_+ H_+ + a_- H_- + b_+ Z_+ + b_- Z_- + c M + d P_5)} \, , 
\end{equation}
where the coefficients $a_{\pm}, b_{\pm}, c, d$ are 
\begin{equation}
a_{\pm} = \pm \frac{z\pm 1}{z \mp 1} \text{sech}\, \tau \, , \qquad
b_{\pm} = \frac{J^2 z}{z^2-1} \cosh \tau \, , \qquad
c = - \tanh \tau \, , \qquad
d= \frac{2 J z}{z^2 - 1}\, . 
\end{equation}
The MAS of $\mathfrak{g}$ is not unique, suppose here we make the choice $\text{MAS} =\{ M, Z, P_2, ..., P_9 \}$.
By using a similarity transformation $S$, we can write (\ref{M_BMN_like}) in the form (\ref{M_non_diag})
\begin{equation}
\mathcal{M} = S e^{2\pi \big(M + \xi Z + d P_5 \big) } S^{-1} \, , 
\end{equation}
where $M, Z, P_5 \in \text{MAS}$, $\xi$ is a coefficient depending on the spectral parameter, 
\begin{equation}\label{xi_coeff}
\xi = - \left(\frac{\sqrt{2} J z}{z^2 - 1}\right)^2 \, ,
\end{equation}
and the similarity transformation is 
\begin{equation}
S = e^{\alpha_+ H_+ + \alpha_- H_- + \beta_+ Z_+ + \beta_- Z_-} \, ,  
\end{equation}
where the coefficients $\alpha_{\pm}, \beta_{\pm}$ are 
\begin{equation}
\begin{aligned}
\alpha_{\pm} &= - \frac{z\pm 1}{z \mp 1} \arccos \left( - \tanh \tau \right) \, , \\
\beta_{\pm} &= \mp \frac{J^2 z}{(z\pm1)(z\mp 1)^3} \bigg[ (z^2 - 1)\arccos\left(-\tanh \tau\right) \\
&\hspace{1cm}\pm 2 z \sinh \tau + (z^2 +1) \arccos\left( - \tanh \tau \right) \sinh^2 \tau \bigg] \, .
\end{aligned}
\end{equation}
Notice that the quasi-momenta accompanying $Z$ and $P_5$ depend on the spectral parameter but the one accompanying $M$ does not. This is because $M$ is the only one of the three generators that is diagonalisable in the adjoint representation, meaning that it is the only one that has to satisfy theorem~\ref{corollary1}.

The generalised quasi-momenta $q_i$ now have a non-trivial dependence on $z$, but does not have any branch cuts in $z$. The simple analytical structure of this solution is similar to that of the BMN string in the relativistic case. More general solutions are expected to have square root branch cuts in the spectral parameter. In the diagonalisable case, the behaviour of the function along those cuts is governed by the \emph{Cartan matrix} of the symmetry group. In the non-diagonalisable case this is no longer applicable, and one needs to find a generalised notion of Cartan matrix acting on the generalised quasi-momenta.

\section{Monodromies for relativistic strings in flat space}\label{sec:flat_space}

In this section we want to temporarily put the non-relativistic strings to one side and make some comments regarding the monodromy matrix eigenvalues for \emph{relativistic} strings in flat space. Our computation is based on choosing the adjoint representation of the generators. 
The flat space action for relativistic strings can be written in the coset form as 
\begin{equation}
S^{\text{flat space}} = - \frac{T}{2} \int \dd^2 \sigma \, \gamma^{\alpha\beta} \langle A^{(1)}_{\alpha} , A^{(1)}_{\beta} \rangle \ , 
\end{equation}
where Minkowski spacetime is written as a coset as $G/H = ISO(1,9) / SO(1,9)$, the $\mathfrak{iso}(1,9)$ algebra is generated by translations $P_a$ and angular momenta $J_{ab}$, and it has a $\mathbb{Z}_2$ outer automorphism, where $P_a$ have grading 1 and $J_{ab}$ grading 0. The inner product is $\mathfrak{so}(1,9)$-adjoint invariant, and can be chosen to be $\langle P_a , P_b \rangle = \eta_{ab}$. A Lax pair is given by 
\begin{equation}
\label{Lax_flat}
\L_\alpha = A_\alpha^{(0)} + \frac{z^2 +1}{z^2-1} A_\alpha^{(1)} - \frac{2z}{z^2-1} \gamma_{\alpha\beta}\varepsilon^{\beta \gamma} A_\gamma^{(1)} \ , 
\end{equation}
and its associated monodromy matrix is constructed with the usual formula (\ref{monodromymatrix}). For the purpose of computing eigenvalues, we take generators in the adjoint representation of $\mathfrak{iso}(1,9)$ following the convention in Appendix \ref{app:Rep}.

As a first step, we compute the eigenvalues of $\L_{\sigma}(z)$, without evaluating it on a particular string solution. It turns out that all  eigenvalues only depend on $\L_{\sigma}^{J_{ab}}$, which by construction do not carry any spectral parameter dependence. Then for a $\L_{\sigma}(z)$ independent of $\sigma$, the eigenvalues of the monodromy matrix also do not depend on $z$. 

In the choice of coordinates given by the coset representative 
\begin{equation}
g = e^{P_a x^a} \ , 
\end{equation}
one can consider the point-like string solution\footnote{It is worthwhile pointing out that the monodromy matrix evaluated on the point-like solution (\ref{point_like_flat}) at large $z$ tends to the identity, due to periodicity of the solution. However, for the NR longitudinal and BMN-like string solutions, $x_1 = \sigma$ is not immediately periodic. The fact $x_1$ is an isometry allows us to compactify $x_1 \sim x_1 + 2\pi$ and interpret $x_1 = \sigma$ as a periodic solution. However, computationally speaking, the periodicity on the coset representative $g(\tau, 2 \pi) = g(\tau, 0)$ does not hold, and therefore formula (\ref{rep_free_asympt_monodromy}) does not apply. } 
\begin{equation}\label{point_like_flat}
t = \kappa \tau \ , \qquad\qquad
x_1 = \kappa \tau \ , \qquad\qquad
\text{others} = 0 \ . 
\end{equation}
The monodromy matrix evaluated on this solution is non-diagonalisable, and already comes in Jordan normal form
\begin{equation}
 \mathcal{M} = e^{i q_0 P_0 + i q_1 P_1} \ , \qquad\qquad
q_0 =  q_1 = \frac{i 4 \pi \kappa z}{1-z^2} \ . 
\end{equation}
Therefore we find once again a monodromy matrix whose eigenvalues are independent of the spectral parameter, but if we express the monodromy matrix as an exponential of generators belonging to a MAS, in this case chosen to be $\{ P_a \}$, then the generalised quasi-momenta $q_i$ retain the spectral parameter dependence. This example shows that the peculiarities that we described in the previous section are not exclusive of non-relativistic strings.

\section{Conclusions}

In this paper we initiated the study of a spectral curve for NR strings in AdS$_5\times S^5$. As a first step, we computed the eigenvalues of the monodromy matrix for a $\sigma$-independent Lax component $\L_{\sigma}(z)$, without evaluating it on any classical string solution. We found that the eigenvalues evaluated just on the constraint surface defined by the Lagrange multipliers equations of motion are always independent of the spectral parameter. 

Then, we evaluated the monodromy matrix on the longitudinal and BMN-like NR string solutions. For the longitudinal solution, the monodromy matrix is diagonalisable and its quasi-momenta are spectral parameter independent, in agreement with the general theorem. In the BMN-like solution, the monodromy matrix is non-diagonalisable. In this case it can be brought into a Jordan normal form, by a similarity transformation. We proposed that the upper triangular form of the non-diagonalisable monodromy matrix in the new basis is realised as an exponential of generators of a maximal abelian subalgebra, and we found a concrete realisation for the BMN-like case. This allowed us to introduce a notion of generalised quasi-momenta, in which case they retain a dependence on the spectral parameter. In particular, for the BMN-like case, their dependence on $z$ is the same as for the relativistic BMN string in AdS$_5\times S^5$. It is interesting to point out that for a generic string solution the generalised quasi-momenta can be $\tau$ dependent, since there are no universal evolution equations that constrain them. 

It remains to understand how to take further the idea of replacing the Cartan subalgebra with the maximal abelian subalgebra at the level of the finite-gap equations, as one needs to replace the Cartan matrix with a different generalised object. In the spectral curve we would expect the sheets of the generalised quasi-momenta to be connected through square root branch cuts. However, the BMN-like solution only gives rise to poles in the spectral curve (as does the BMN string in AdS$_5 \times S^5$). 
It would be useful to consider the monodromy matrix of the NR analogue of GKP solution \cite{Fontanella:2021btt}, since the corresponding solution would give rise to branch cuts in the relativistic string.

Since the eigenvalues of the monodromy matrix are independent of the spectral parameter, it is not possible to directly extract higher conserved charges from the transfer matrix. However, this does not automatically mean that the model is not integrable. Indeed, as we have seen, the transfer matrix of the relativistic string in flat space also does not generate any higher conserved charges, but the free field theory of the flat space string gives rise to many additional conservation laws. It would be interesting to further investigate the integrability properties of the NR string and find a generalisation of the transfer matrix which encodes any possible higher charges of the theory.

Finally it would be interesting to understand the relativistic origin of what we call BMN-like NR string, in the sense whether there exists a relativistic string solution that can be connected by a limit to our BMN-like solution. In such case, it would be interesting to understand if one can take a limit in the monodromy matrix associated with such relativistic string solution and recover the result found here for the BMN-like solution. Since in the relativistic theory the monodromy matrix is diagonalisable, it would be instructive to understand where the diagonalisability property breaks down in the limit.

\section*{Acknowledgments}

We are grateful to P.~Vieira for useful discussions and to B.~Stefa\'nski for useful comments on a draft of this work. Research at Perimeter Institute is supported in part by the Government of Canada through the Department of Innovation, Science and Economic Development and by the Province of Ontario through the Ministry of Colleges and Universities. AF thanks the Department of Physics and Astronomy at the University of Padova for hospitality and for financially supporting his visit during the completion of part of this work. JMNG is supported by the EPSRC-SFI grant EP/S020888/1 \emph{Solving Spins and Strings}. The work of OOS was supported by VR grant 2021-04578.  Nordita is supported in part by NordForsk.
AF thanks Lia for her permanent support.


\setcounter{section}{0}
\setcounter{subsection}{0}
\setcounter{equation}{0}

\begin{appendices}

\section{Conventions}
\label{app:Conventions}

For a generic object $\mathcal{O}^A$, we define its light-cone combinations as
\begin{equation}\label{LC_comb}
\mathcal{O}^{\pm} \equiv \mathcal{O}^0 \pm \mathcal{O}^1 \ , \qquad\qquad
\mathcal{O}_{\pm} \equiv \frac{1}{2}\left( \mathcal{O}_0 \pm \mathcal{O}_1 \right) \ .
\end{equation}
The longitudinal Minkowski metric then has non-vanishing components $\eta_{+-} = -1/2$ and $\eta^{+-} = -2$. We take $\varepsilon^{01} = - \varepsilon_{01} = + 1$ for $\varepsilon^{\alpha\beta}$, $\varepsilon^{\sf ab}$ and $\varepsilon^{AB}$. In light-cone components $\varepsilon_{+-} = \frac{1}{2}$, $\varepsilon^{+-} = -2$.

The Hodge dual of a $p$-form $\omega = \frac{1}{p!} \omega_{\mu_1 \cdots \mu_p} \dd x^{\mu_1}\wedge \cdots \wedge \dd x^{\mu_p}$ is
\begin{equation}
\star \omega = \frac{\sqrt{|g|}}{p!(D-p)!} \omega_{\mu_1 \cdots \mu_p} \varepsilon^{\mu_1 \cdots \mu_p}{}_{\nu_{p+1} \cdots \nu_D}  \dd x^{\nu_{p+1}}\wedge \cdots \wedge \dd x^{\nu_D} \ .
\end{equation}

\section{Representations and inner products}
\label{app:Rep}

\noindent {\bf Spinorial representation.} The spinorial representation of the algebra (\ref{ext_NH_algebra}), (\ref{Euclidean}) has been constructed in \cite{Fontanella:2022pbm}, and it is inherited by Lie algebra expansion from the spinorial representation of $\mathfrak{so}(2,4) \oplus \mathfrak{so}(6)$. Here we shall review it. We split the generators into three families created by the particular Lie algebra expansion in \cite{Fontanella:2020eje}, i.e., $\mathfrak{g} = \mathfrak{g}_0 \oplus \mathfrak{g}_1 \oplus \mathfrak{g}_2$,
\begin{equation}
\begin{aligned}
\mathfrak{g}_0 &= \{M, H_A, J_{ab}, J_{a'b'} \} \, , \\
\mathfrak{g}_1 &= \{G_{Aa}, P_a, P_{a'} \} \, , \\
\mathfrak{g}_2 &= \{Z, Z_A, Z_{ab},  Z_{a'b'} \} \, . 
\end{aligned}
\end{equation}
Each family creates a \emph{level}, and to each level we associate a matrix $\omega_i$, with $i = 0,1,2$, given by 
\begin{equation}
\omega_0 = \begin{pmatrix}
1 & 0 & 0\\
0 & 1 & 0 \\
0 & 0 & 1
\end{pmatrix},  \quad 
\omega_1 = \begin{pmatrix}
0 & 0 & 0\\
1 & 0 & 0 \\
0 & 1 & 0
\end{pmatrix} , \quad
\omega_2 = \begin{pmatrix}
0 & 0 & 0\\
0 & 0 & 0 \\
1 & 0 & 0
\end{pmatrix} \ . 
\end{equation}
Then for a generator $\hat{X} \in \mathfrak{g}_i$ of level $i$, its spinorial representation $\hat{\rho}$ is 
\begin{equation}
\hat{\rho}(\hat{X}) \equiv \omega_i \otimes \rho ( X ) \ ,
\end{equation}
where $\rho ( X )$ is the spinorial representation of the parental generator in $\mathfrak{so}(2,4) \oplus \mathfrak{so}(6)$ from which $\hat{X}$ comes from, accordingly to the following Lie algebra expansion rule
\begin{equation}\label{LAE}
\begin{aligned}
J_{AB} &\rightarrow \varepsilon_{AB} ( M + \epsilon^2 Z ) \, ,\\
J_{Aa} &\rightarrow \epsilon G_{Aa}\, ,\\
J_{ab} &\rightarrow J_{ab} + \epsilon^2 Z_{ab} \, , \\
P_{A}&\rightarrow H_A + \epsilon^2 Z_A \, , \\
P_{a} &\rightarrow \epsilon P_a \, ,\\
J_{a'b'}&\rightarrow J_{a'b'} + \epsilon^2 Z_{a'b'}\, , \\
P_{a'}&\rightarrow \epsilon P_{a'} \, . 
\end{aligned}
\end{equation}
Our convention for the spinorial representation of generators of $\mathfrak{so}(2,4) \oplus \mathfrak{so}(6)$ is the same as in \cite{Arutyunov:2009ga}. A $\mathfrak{g}$-adjoint invariant  inner product for the spinorial representation can be constructed with the general form 
\begin{equation}
\langle \hat{\rho}(\hat{X}), \hat{\rho}(\hat{Y}) \rangle \equiv \text{STr} \left[\left( (a \,\omega_0 +  b \, \omega_1 + c\, \omega_2)^t \otimes \mathbf{1}\right) \hat{\rho}(\hat{X}) \hat{\rho}(\hat{Y} )\right] \, , 
\end{equation}
where $a, b, c$ constants. We need to set $a=0$ in order to have $\langle H_+ , H_- \rangle = 0$, as required by the string Newton-Cartan structure for the NR string. The supertrace `STr' is defined as 
\begin{equation}
\text{STr} (m \otimes \mathcal{M}) \equiv \text{Tr}(m) \text{STr}(\mathcal{M}) \ , 
\end{equation}
where the supertrace in the spinorial representation is computed in the usual way, i.e., $\text{STr}(\mathcal{M}) = \text{Tr}_{AdS} (\mathcal{M}) - \text{Tr}_{S}(\mathcal{M})$.

\vspace{2mm}
\noindent {\bf Adjoint representation.} Another independent check about the eigenvalues of the monodromy matrix come from considering the adjoint representation of the initial (equivalent) model given in \cite{Fontanella:2022fjd}. The algebra is less extended than in \cite{Fontanella:2022pbm}, and it has less nice properties. Nevertheless it provides an independent check of the result.    

Our convention for the adjoint representation is 
\begin{equation}
\big( \ad_{X_a} \big)_b{}^c = - f_{ab}{}^c  \, , \qquad\qquad\forall \ X_a \in \mathfrak{g} \, ,
\end{equation}
where $f_{ab}{}^c$ are the structure constants of $\mathfrak{g}$ given in \cite{Fontanella:2022fjd}, formula (1.4). This representation is faithful and non-unitary. 
To construct an inner product, it is useful to note that the matrix product $(\ad_{X_a})^T \ad_{X_b}$ is non-zero only if $a=b$. This allows us to introduce a non-degenerate, $\mathbb{Z}_2$ grading compatible, inner product $\langle \cdot, \cdot \rangle_{\text{ad rep}}$ between generators in the adjoint representation, 
\begin{equation}
\langle\ad_{X_a}, \ad_{X_b} \rangle_{\text{ad rep}} \equiv \frac{(\ad_{X_a})^T \ad_{X_b}}{(\ad_{X_a})^T\ad_{X_a}}  = \delta_{ab} \ . 
\end{equation}
Such inner product is not invariant under the adjoint action of $\tilde{\mathfrak{h}}\equiv \mathfrak{h} \setminus \{ Z_A \}$ as demanded in \cite{Fontanella:2022fjd} because, e.g., $\langle\ad_{Z_+}, \ad_{Z_+} \rangle_{\text{rep}}$ is non-zero. However for the purpose of extracting components from a MC 1-form, this inner product is enough.  One can use such inner product to define projectors into $\mathfrak{g}^{(0)}$ and $\mathfrak{g}^{(1)}$ of a generic algebra-valued function $\Theta$ in the adjoint representation, as
\begin{equation}
\Theta^{(0)} = \sum_{X_a \in\, \mathfrak{g}^{(0)}}  \left[\frac{(\ad_{X_a})^T \Theta}{(\ad_{X_a})^T\ad_{X_a}}\right] \ad_{X_a} \, , \qquad\quad
\Theta^{(1)} = \sum_{X_a \in\, \mathfrak{g}^{(1)}}  \left[\frac{(\ad_{X_a})^T \Theta}{(\ad_{X_a})^T\ad_{X_a}}\right] \ad_{X_a}  \, .
\end{equation}
The generators whose adjoint representation is diagonalisable are
\begin{equation}
H_0, \qquad H_1, \qquad M, \qquad J_{ab}, \qquad J_{a'b'} \, , 
\end{equation}  
Notice that $H_{\pm}$ are not diagonalisable. Cartan subalgebras are constructed by picking one generator among $\{H_0, H_1, M\}$ together with a maximal set of commuting angular momentum generators, e.g., 
\begin{equation}
\{ H_0, J_{23}, J_{56}, J_{78}\} \, , \qquad
\{ H_1, J_{23}, J_{56}, J_{78}\} \, , \qquad
\{ M, J_{23}, J_{56}, J_{78}\} \, . 
\end{equation}

\end{appendices}


\bibliographystyle{nb}

\bibliography{Name}

\begin{thebibliography}{10}
\ifx\href\asklfhas\newcommand{\href}[2]{#2}\fi
\ifx\arxivref\asklfhas\newcommand{\arxivref}[2]{\href{http://arxiv.org/abs/#1}{#2}}\fi
\ifx\doiref\asklfhas\newcommand{\doiref}[2]{\href{http://dx.doi.org/#1}{#2}}\fi
\raggedright
\small
\parskip 0pt

\bibitem{Gomis:2005pg}
J.~Gomis, J.~Gomis and K.~Kamimura,
\textit{``{Non-relativistic superstrings: A New soluble sector of AdS(5) x
  S**5}''},
\textsf{\doiref{10.1088/1126-6708/2005/12/024}{JHEP~0512,~024~(2005)}},
\texttt{\arxivref{hep-th/0507036}{hep-th/0507036}}.

\bibitem{Harmark:2017rpg}
T.~Harmark, J.~Hartong and N.~A.~Obers,
\textit{``{Nonrelativistic strings and limits of the AdS/CFT
  correspondence}''},
\textsf{\doiref{10.1103/PhysRevD.96.086019}{Phys.~Rev.~D~96,~086019~(2017)}},
\texttt{\arxivref{1705.03535}{arxiv:1705.03535}}.

\bibitem{Harmark:2018cdl}
T.~Harmark, J.~Hartong, L.~Menculini, N.~A.~Obers and Z.~Yan,
\textit{``{Strings with Non-Relativistic Conformal Symmetry and Limits of the
  AdS/CFT Correspondence}''},
\textsf{\doiref{10.1007/JHEP11(2018)190}{JHEP~1811,~190~(2018)}},
\texttt{\arxivref{1810.05560}{arxiv:1810.05560}}.

\bibitem{Harmark:2019upf}
T.~Harmark, J.~Hartong, L.~Menculini, N.~A.~Obers and G.~Oling,
\textit{``{Relating non-relativistic string theories}''},
\textsf{\doiref{10.1007/JHEP11(2019)071}{JHEP~1911,~071~(2019)}},
\texttt{\arxivref{1907.01663}{arxiv:1907.01663}}.

\bibitem{Gomis:2000bd}
J.~Gomis and H.~Ooguri,
\textit{``{Nonrelativistic closed string theory}''},
\textsf{\doiref{10.1063/1.1372697}{J.~Math.~Phys.~42,~3127~(2001)}},
\texttt{\arxivref{hep-th/0009181}{hep-th/0009181}}.

\bibitem{Danielsson:2000gi}
U.~H.~Danielsson, A.~Guijosa and M.~Kruczenski,
\textit{``{IIA/B, wound and wrapped}''},
\textsf{\doiref{10.1088/1126-6708/2000/10/020}{JHEP~0010,~020~(2000)}},
\texttt{\arxivref{hep-th/0009182}{hep-th/0009182}}.

\bibitem{Bergshoeff:2018yvt}
E.~Bergshoeff, J.~Gomis and Z.~Yan,
\textit{``{Nonrelativistic String Theory and T-Duality}''},
\textsf{\doiref{10.1007/JHEP11(2018)133}{JHEP~1811,~133~(2018)}},
\texttt{\arxivref{1806.06071}{arxiv:1806.06071}}.

\bibitem{Gomis:2019zyu}
J.~Gomis, J.~Oh and Z.~Yan,
\textit{``{Nonrelativistic String Theory in Background Fields}''},
\textsf{\doiref{10.1007/JHEP10(2019)101}{JHEP~1910,~101~(2019)}},
\texttt{\arxivref{1905.07315}{arxiv:1905.07315}}.

\bibitem{Bergshoeff:2019pij}
E.~A.~Bergshoeff, J.~Gomis, J.~Rosseel, C.~\c{S}im\c{s}ek and Z.~Yan,
\textit{``{String Theory and String Newton-Cartan Geometry}''},
\textsf{\doiref{10.1088/1751-8121/ab56e9}{J.~Phys.~A~53,~014001~(2020)}},
\texttt{\arxivref{1907.10668}{arxiv:1907.10668}}.

\bibitem{Hartong:2021ekg}
J.~Hartong and E.~Have,
\textit{``{Nonrelativistic Expansion of Closed Bosonic Strings}''},
\textsf{\doiref{10.1103/PhysRevLett.128.021602}{Phys.~Rev.~Lett.~128,~021602~(2022)}},
\texttt{\arxivref{2107.00023}{arxiv:2107.00023}}.

\bibitem{Hartong:2022dsx}
J.~Hartong and E.~Have,
\textit{``{Nonrelativistic Approximations of Closed Bosonic String Theory}''},
\texttt{\arxivref{2211.01795}{arxiv:2211.01795}}.

\bibitem{Bidussi:2021ujm}
L.~Bidussi, T.~Harmark, J.~Hartong, N.~A.~Obers and G.~Oling,
\textit{``{Torsional string Newton-Cartan geometry for non-relativistic
  strings}''},
\textsf{\doiref{10.1007/JHEP02(2022)116}{JHEP~2202,~116~(2022)}},
\texttt{\arxivref{2107.00642}{arxiv:2107.00642}}.

\bibitem{Ko:2015rha}
S.~M.~Ko, C.~Melby-Thompson, R.~Meyer and J.-H.~Park,
\textit{``{Dynamics of Perturbations in Double Field Theory \& Non-Relativistic
  String Theory}''},
\textsf{\doiref{10.1007/JHEP12(2015)144}{JHEP~1512,~144~(2015)}},
\texttt{\arxivref{1508.01121}{arxiv:1508.01121}}.

\bibitem{Blair:2019qwi}
C.~D.~A.~Blair,
\textit{``{A worldsheet supersymmetric Newton-Cartan string}''},
\textsf{\doiref{10.1007/JHEP10(2019)266}{JHEP~1910,~266~(2019)}},
\texttt{\arxivref{1908.00074}{arxiv:1908.00074}}.

\bibitem{Blair:2020gng}
C.~D.~A.~Blair, G.~Oling and J.-H.~Park,
\textit{``{Non-Riemannian isometries from double field theory}''},
\textsf{\doiref{10.1007/JHEP04(2021)072}{JHEP~2104,~072~(2021)}},
\texttt{\arxivref{2012.07766}{arxiv:2012.07766}}.

\bibitem{Blair:2020ops}
C.~D.~A.~Blair,
\textit{``{Non-relativistic duality and $T \bar T$ deformations}''},
\textsf{\doiref{10.1007/JHEP07(2020)069}{JHEP~2007,~069~(2020)}},
\texttt{\arxivref{2002.12413}{arxiv:2002.12413}}.

\bibitem{Kluson:2017abm}
J.~Kluson,
\textit{``{Note about Hamiltonian formalism for Newton\textendash{}Cartan
  string and p-brane}''},
\textsf{\doiref{10.1140/epjc/s10052-018-5993-8}{Eur.~Phys.~J.~C~78,~511~(2018)}},
\texttt{\arxivref{1712.07430}{arxiv:1712.07430}}.

\bibitem{Kluson:2018egd}
J.~Kluso\v{n},
\textit{``{Remark About Non-Relativistic String in Newton-Cartan Background and
  Null Reduction}''},
\textsf{\doiref{10.1007/JHEP05(2018)041}{JHEP~1805,~041~(2018)}},
\texttt{\arxivref{1803.07336}{arxiv:1803.07336}}.

\bibitem{Kluson:2018grx}
J.~Kluso\v{n},
\textit{``{Nonrelativistic String Theory Sigma Model and Its Canonical
  Formulation}''},
\textsf{\doiref{10.1140/epjc/s10052-019-6623-9}{Eur.~Phys.~J.~C~79,~108~(2019)}},
\texttt{\arxivref{1809.10411}{arxiv:1809.10411}}.

\bibitem{Gomis:2020fui}
J.~Gomis, Z.~Yan and M.~Yu,
\textit{``{Nonrelativistic Open String and Yang-Mills Theory}''},
\textsf{\doiref{10.1007/JHEP03(2021)269}{JHEP~2103,~269~(2021)}},
\texttt{\arxivref{2007.01886}{arxiv:2007.01886}}.

\bibitem{Gomis:2020izd}
J.~Gomis, Z.~Yan and M.~Yu,
\textit{``{T-Duality in Nonrelativistic Open String Theory}''},
\textsf{\doiref{10.1007/JHEP02(2021)087}{JHEP~2102,~087~(2021)}},
\texttt{\arxivref{2008.05493}{arxiv:2008.05493}}.

\bibitem{Oling:2022fft}
G.~Oling and Z.~Yan,
\textit{``{Aspects of Nonrelativistic Strings}''},
\texttt{\arxivref{2202.12698}{arxiv:2202.12698}}.

\bibitem{Fontanella:2021btt}
A.~Fontanella and J.~M.~N.~Garc\'\i{}a,
\textit{``{Classical string solutions in non-relativistic AdS$_{5}$
  \texttimes{} S$^{5}$: closed and twisted sectors}''},
\textsf{\doiref{10.1088/1751-8121/ac4abd}{J.~Phys.~A~55,~085401~(2022)}},
\texttt{\arxivref{2109.13240}{arxiv:2109.13240}}.

\bibitem{Fontanella:2021hcb}
A.~Fontanella, J.~M.~Nieto~Garc\'\i{}a and A.~Torrielli,
\textit{``{Light-Cone Gauge in Non-Relativistic AdS$_5\times$S$^5$ String
  Theory}''},
\texttt{\arxivref{2102.00008}{arxiv:2102.00008}}.

\bibitem{Fontanella:2022fjd}
A.~Fontanella and S.~J.~van~Tongeren,
\textit{``{Coset space actions for nonrelativistic strings}''},
\textsf{\doiref{10.1007/JHEP06(2022)080}{JHEP~2206,~080~(2022)}},
\texttt{\arxivref{2203.07386}{arxiv:2203.07386}}.

\bibitem{Fontanella:2022pbm}
A.~Fontanella and J.~M.~Nieto~Garc\'\i{}a,
\textit{``{Extending the non-relativistic string AdS coset}''},
\texttt{\arxivref{2208.02295}{arxiv:2208.02295}}.

\bibitem{Arutyunov:2009ga}
G.~Arutyunov and S.~Frolov,
\textit{``{Foundations of the AdS$_5\times$S$^5$ Superstring. Part I}''},
\textsf{\doiref{10.1088/1751-8113/42/25/254003}{J.Phys.~A42,~254003~(2009)}},
\texttt{\arxivref{0901.4937}{arxiv:0901.4937}}.

\bibitem{Beisert:2010jr}
N.~Beisert et~al.,
\textit{``Review of {AdS/CFT} Integrability: An Overview''},
\textsf{\doiref{10.1007/s11005-011-0529-2}{Lett.~Math.~Phys.~99,~3~(2012)}},
\texttt{\arxivref{1012.3982}{arxiv:1012.3982}}.

\bibitem{Stefanski:2008ik}
B.~Stefa{\'n}ski,~jr,
\textit{``{G}reen-{S}chwarz action for Type {IIA} strings on {AdS$_4\times
  CP^3$}''},
\textsf{\doiref{10.1016/j.nuclphysb.2008.09.015}{Nucl.~Phys.~B808,~80~(2009)}},
\texttt{\arxivref{0806.4948}{arxiv:0806.4948}}.

\bibitem{Arutyunov:2008if}
G.~Arutyunov and S.~Frolov,
\textit{``Superstrings on {AdS$_4 \times CP^3$} as a Coset Sigma-model''},
\textsf{\doiref{10.1088/1126-6708/2008/09/129}{JHEP~0809,~129~(2008)}},
\texttt{\arxivref{0806.4940}{arxiv:0806.4940}}.

\bibitem{Babichenko:2009dk}
A.~Babichenko, B.~Stefa{\'n}ski,~jr. and K.~Zarembo,
\textit{``Integrability and the {AdS${}_{3}$/CFT${}_{2}$} correspondence''},
\textsf{\doiref{10.1007/JHEP03(2010)058}{JHEP~1003,~058~(2010)}},
\texttt{\arxivref{0912.1723}{arxiv:0912.1723}}.

\bibitem{OhlssonSax:2011ms}
O.~Ohlsson~Sax and B.~Stefa{\'n}ski,~jr.,
\textit{``Integrability, spin-chains and the {AdS${}_{3}$/CFT${}_{2}$}
  correspondence''},
\textsf{\doiref{10.1007/JHEP08(2011)029}{JHEP~1108,~029~(2011)}},
\texttt{\arxivref{1106.2558}{arxiv:1106.2558}}.

\bibitem{Cagnazzo:2012se}
A.~Cagnazzo and K.~Zarembo,
\textit{``{B-field in AdS(3)/CFT(2) Correspondence and Integrability}''},
\textsf{\doiref{10.1007/JHEP11(2012)133}{JHEP~1211,~133~(2012)}},
\texttt{\arxivref{1209.4049}{arxiv:1209.4049}},
[Erratum: JHEP 04, 003 (2013)].

\bibitem{SchaferNameki:2010jy}
S.~Sch{\"a}fer-Nameki,
\textit{``Review of {AdS/CFT} Integrability, {C}hapter {II.4}: The Spectral
  Curve''},
\textsf{\doiref{10.1007/s11005-011-0525-6}{Lett.~Math.~Phys.~99,~169~(2010)}},
\texttt{\arxivref{1012.3989}{arxiv:1012.3989}}.

\bibitem{Zarembo:2010yz}
K.~Zarembo,
\textit{``Algebraic Curves for Integrable String Backgrounds''},
\texttt{\arxivref{1005.1342}{arxiv:1005.1342}}.

\bibitem{Gromov:2008ec}
N.~Gromov, S.~Schafer-Nameki and P.~Vieira,
\textit{``{Efficient precision quantization in AdS/CFT}''},
\textsf{\doiref{10.1088/1126-6708/2008/12/013}{JHEP~0812,~013~(2008)}},
\texttt{\arxivref{0807.4752}{arxiv:0807.4752}}.

\bibitem{Fontanella:2020eje}
A.~Fontanella and L.~Romano,
\textit{``{Lie Algebra Expansion and Integrability in Superstring
  Sigma-Models}''},
\textsf{\doiref{10.1007/JHEP07(2020)083}{JHEP~2007,~083~(2020)}},
\texttt{\arxivref{2005.01736}{arxiv:2005.01736}}.

\bibitem{Gromov:2007aq}
N.~Gromov and P.~Vieira,
\textit{``{The $AdS_5 \times S^5$ superstring quantum spectrum from the
  algebraic curve}''},
\textsf{\doiref{10.1016/j.nuclphysb.2007.07.032}{Nucl.~Phys.~B~789,~175~(2008)}},
\texttt{\arxivref{hep-th/0703191}{hep-th/0703191}}.

\bibitem{Gromov:2007cd}
N.~Gromov and P.~Vieira,
\textit{``{Constructing the AdS/CFT dressing factor}''},
\textsf{\doiref{10.1016/j.nuclphysb.2007.08.019}{Nucl.~Phys.~B~790,~72~(2008)}},
\texttt{\arxivref{hep-th/0703266}{hep-th/0703266}}.

\bibitem{Gromov:2008ie}
N.~Gromov, S.~Schafer-Nameki and P.~Vieira,
\textit{``{Quantum Wrapped Giant Magnon}''},
\textsf{\doiref{10.1103/PhysRevD.78.026006}{Phys.~Rev.~D~78,~026006~(2008)}},
\texttt{\arxivref{0801.3671}{arxiv:0801.3671}}.

\end{thebibliography}

\end{document}